\documentclass[]{aa}
\usepackage{psfig}
\usepackage{times}
\usepackage{graphics}
\begin{document}

\title{Research Note\\
The EUV variability of the luminous QSO HS~1700+6416}
\author {D. Reimers \inst{1}
\and H.-J. Hagen \inst{1}
\and J. Schramm \inst{1}
\and G.A. Kriss, \inst{2}
\and J.M. Shull \inst{3}}

\institute{Hamburger Sternwarte, Universit\"at Hamburg, Gojenbergsweg 112, D-21029 Hamburg, Germany
\and STScI, 3700 San Martin Dr. Baltimore, MD 21218, USA
\and CASA, Department of Astrophysical and Planetary Science, University of
Colorado, Boulder, CO 80309}
\offprints{D.Reimers, dreimers@hs.uni-hamburg.de}
\date{received date; accepted date}

\abstract{We report on the first observations of variations in UV
(intrinsic EUV 330 \AA\,) flux of the luminous QSO HS~1700+6416 (z =
2.72) over a decade. The amplitude of variations increases from $\pm$
0.1 mag in the optical (R) to up to a factor of 3 at 1250 \AA\,.  This
is apparently an extension of the increase in amplitude of
variations towards shorter wavelengths observed with IUE in low z AGN
(Paltani \& Courvoisier, 1996) to the EUV. The time-scale for
variations with the largest amplitudes is $\geq$ 1/2 yr to years. We
 briefly discuss the consequences of the observed variations on the
ionizing metagalactic UV background.

\keywords{galaxies: quasars: individual HS~1700+6416 -- galaxies: quasars: general}
}
\maketitle
\markboth{D. Reimers et al.: The EUV variability of the luminous QSO HS~1700+6416}{}
\section{Introduction}
The origin of the UV spectral energy distribution of active galactic
nuclei (AGN) and, in particular, that of luminous QSOs is poorly
understood. Typically, QSOs show a steep power law type spectrum
$f_{\nu}\sim\nu^{-\alpha}$ with $<\alpha>$ = 1.8 (e.g., Telfer et
al. \cite{tel}). It is also known from the compilation of QSO UV flux
distributions from the HST archive (Telfer et al.  \cite{tel}) and from
the FUSE archive (Scott et al. \cite{sco04}) that
the UV spectral index shows a broad distribution ($0 \leq \alpha \leq
3$). The four luminous QSOs, for which the energy distribution has been
observed at intrinsic EUV wavelengths as short as 300 {\AA}, are all
individuals, and the spectral distribution of the so-called "big blue
bump" cannot be modelled by simple power laws (cf. Fig. 2 in Reimers
et al. \cite{rei98}). Little is known about the intrinsic UV/EUV
variability of luminous QSOs. In a systematic study of the UV
variability of AGN using the IUE archive, Paltani \& Courvoisier (\cite{pal94})
 found that most AGN show flux variations whose amplitude
increases from rest wavelengths 3200 \AA\, to 1200 \AA\, by roughly a
factor of 2.  Little is known about simultaneous optical variations
of AGN, except in a few cases like 3C~273, NGC~4151, and NGC~5548 (see below).
In this paper
we report on UV (intrinsic EUV) and optical variability of
HS~1700+6416 (z = 2.72), one of the most luminous QSOs known and also
one of the few known objects where such an empirical study is
possible. Discovered by the Hamburg Quasar Survey it was observed
in the UV by IUE (Reimers, \cite{rei89}), HST (Reimers et al.,
\cite{rei92}), HUT (Davidsen et al. \cite{dav}), and FUSE. It was also
 monitored in the optical in the years 1988 - 1995 and 1998 by the
Hamburg Quasar Monitoring Program (Borgeest \& Schramm, \cite{bor})
and from 1995 on at the Wise observatory (Kaspi et al. \cite{kas}).
While the variability of HS~1700+6416 was therefore known before, also
in the UV (cf. discussion by K\"ohler et al.  \cite{koh}),
motivation for the present study came from FUSE observations taken in
Feb/March 2003 when it was recognized that HS~1700+6416 had
brightened by a factor of $\sim$ 3 since Oct. 2002 at FUSE wavelengths.
With EUV rest wavelength
 observations available now at 9 epochs between 1988 and
2003 and optical R-band CCD photometry (rest wavelength 1900 \AA\,)
over several years, we can investigate the relation between the small
amplitude ($\pm$ 10 \%) flux variations in the optical and the large
amplitude variations in the intrinsic EUV.

\begin{figure}
  \centering
  \resizebox{\hsize}{!}{\includegraphics{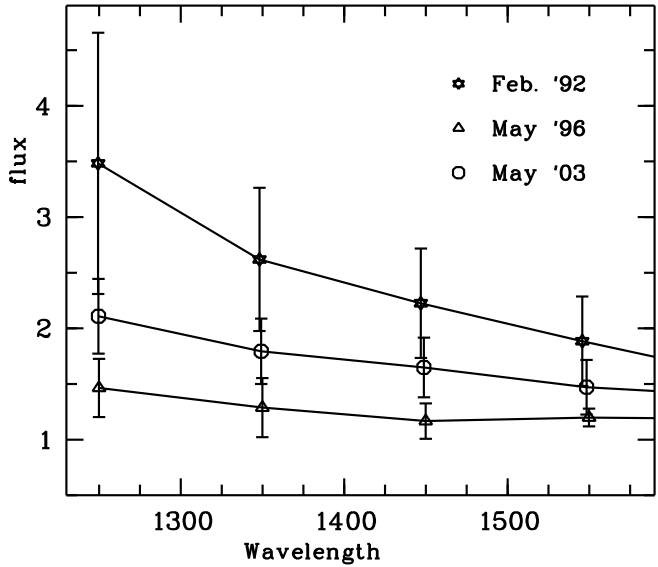}} 
\caption
{
Ultraviolet energy distribution of HS~1700+6416 
 [10$^{-15}$ erg s$^{-1}$ cm$^{-2}$ \AA$^{-1}$] for  selected epochs (cf. Table 1).
}
\label{hstspec}
\end{figure}

\section{Observations}
\subsection{UV data}
We collected all available UV observations of HS~1700+6416
from the IUE and HST archives, most taken by ourselves. In
addition, new specific UV observations were made with STIS
onboard HST in May 2003 with the aim determining the QSO
continuum for interpreting the FUSE intergalactic
HeII 304 \AA\, absorption spectrum. These data will be described
in detail in a later paper.
In Table 1 we present an overview of the UV data collected
between 1988 and 2003. Flux distributions are shown in Fig. \ref{hstspec}
In order to minimize the noise, QSO mean fluxes were formed over 100 \AA\,
 intervals. As expected, and briefly discussed already by K\"ohler, Reimers,
\& Wamsteker (\cite{koh}), the amplitude of the flux variation increases
to shorter wavelengths and reaches a factor of $\geq$ 3 at 1250 \AA\,, while at $\sim$ 1600 \AA\, the factor is less than 2.

\begin{figure}
  \centering
  \resizebox{\hsize}{!}{\includegraphics{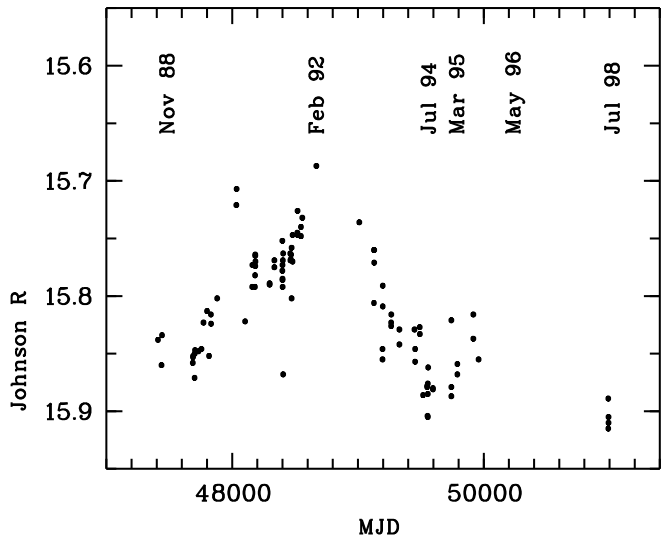}} 
\caption
{ Light curve of HS~1700+6416 from the Hamburg Quasar
monitoring program (Borgeest \& Schramm, \cite{bor}). The dates of (quasi)
simultaneous UV observations are indicated.
}
\label{hqmdat}
\end{figure}

\subsection{Optical monitoring}
HS~1700+6416 was monitored photometrically between the end of 1988 and
mid 1995 by the Hamburg Quasar monitoring program (Borgeest \&
Schramm, \cite{bor}) in the Johnson R-band using the Calar Alto 1.23m
telescope. All available HS~1700+6416 data from the HQM monitoring program
are shown in Fig. \ref{hqmdat}. HS~1700+6416 varies erratically on
time-scales of several months to a few years with a full amplitude
of $\bigtriangleup$ R = 0.2 mag. Variations within a few days appear
to be $\leq$ 0.02 mag and within a month typically less than 0.05 mag. We
mention these maximum amplitudes on short time-scales explicitly, since
while our UV data from space and the data from optical CCD photometry
are not always strictly simultaneous, we wish to use the closest (in
time) optical measurement for a comparison with space data. The
typical rise time to maximum brightness (May 90 and Feb 92) is $\sim$
6 months. 

The optical brightness of HS~1700+6416 for May 2003 was derived from the 
STIS target acquisition exposures.

\subsection{X-ray observations}
Does the variability continue to even shorter wavelengths? ROSAT
observations of HS~1700+6416 at 3 epochs (RASS, and 2 epochs
observed by Reimers et al. \cite{rei97}) show variations apparently out
of phase with the optical data. However, since no strictly
simultaneous X-ray and optical observations are available, no
safe conclusions are possible, except that HS~1700+6416 varies also in 
the ROSAT - band by nearly a factor of $\sim$ 2.
The X-ray flux varies on much shorter time-scales. A 16 ksec
observation with the ROSAT PSPC on Nov 13, 1992, distributed over $\sim$ 21 hours
 shows that the flux varied by a factor of $\sim$ 2 within a day,
so that  no relation between X-ray and optical / UV fluxes
could be established.

\begin{figure}
  \centering
  \resizebox{\hsize}{!}{\includegraphics{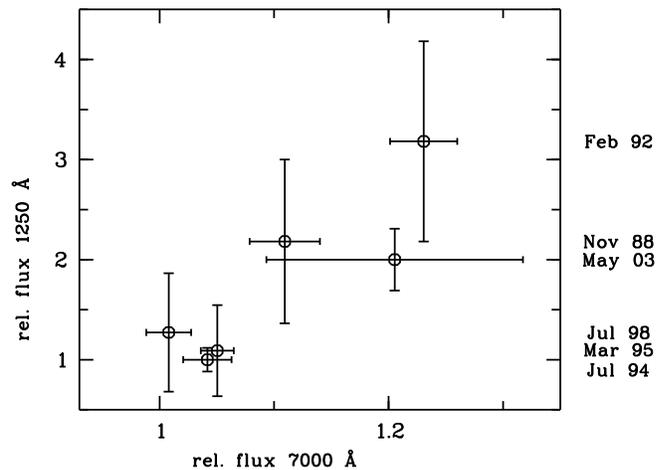}} 
\caption
{
Flux at 1250 \AA\,(in units of 1.1 $\cdot$ 10$^{-15}$ erg s$^{-1}$ cm$^{-2}$ \AA$^{-1}$)
 versus
R band flux (in units of 7.5 $\cdot$ 10$^{-16}$ erg s$^{-1}$ cm$^{-2}$ \AA$^{-1}$).
}
\label{fluxcomp}
\end{figure}

\section{Interpretation and discussion}
The question is whether luminous QSOs behave like Sy1 galaxies in how the
amplitude of variability increases towards shorter wavelengths.  In
Fig. \ref{hqmdat} we marked the  epochs of UV observations
along the R-light curve of HS~1700+6416.  While we have barely any
strictly simultaneous observations, we notice that when the QSO is
bright in R (Feb. 92), it is bright at 1250 \AA\, and vice versa. The
close relation between the flux at 1250 \AA\, ($\lambda_{rest} \simeq$
335 \AA\,) and the flux in the Johnson R-band ($\lambda_{rest}$ = 1900 \AA\,)
is shown in Fig. \ref{fluxcomp}. Although the optical CCD photometry
in R is typically only within a month (except the maximum on Feb. 17,
1992 where we have a simultaneous measurement) the tendency is
clear. An amplitude of $\sim$ 20 \% at $\lambda_{rest}$ = 1900 \AA\,
steepens to an amplitude of a factor of $\sim$ 1.8 at $\lambda_{rest}$
= 420 \AA\, and a factor of $\sim$ 3 at $\lambda_{rest}$ = 330
\AA\,. At face value this looks like a continuation of the trend known
from Sy1 nuclei (Paltani \& Courvoisier, \cite{pal94}) where the amplitude of
variations increases by typically a factor of 2 between
$\lambda_{rest}$ = 3200 \AA\, and $\lambda_{rest}$ = 1200 \AA\,. A
detailed variability study of 25000 SDSS quasars by Vanden Berk et
al. (\cite{van}) has also shown that quasars are about twice as variable at
1000 \AA\, as at 6000 \AA\,.

      The behaviour of HS~1700+6416 at intrinsic EUV wavelengths is
      similar to the EUV behaviour of the few Sy 1 AGN observed so far,
      but it is not as variable at longer wavelengths. In NGC~5548 the
      amplitude of variations increases from visible wavelengths
      through the ultraviolet to the extreme ultraviolet. In the
      visible, variations have an amplitude of $\sim50$\% from maximum
      to minimum, increasing to about a factor of 2 in the
      far-ultraviolet (Korista et al. \cite{kor}), which is much larger than
      the variability seen in HS~1700+6416 at comparable rest
      wavelengths. EUVE observations of NGC~5548 at rest wavelengths
      of $\sim 80$ \AA\ also show factor of 2 variations (Marshall et
      al. \cite{mar97}; Chiang et al. \cite{chi}). The EUV variability of the
      narrow-line Seyfert 1 galaxy Mrk 478 has a comparably large
      amplitude (Marshall et al. \cite{mar96}).

      The current view of UV and EUV emission in AGN is that the broad
      peak of emission in the UV, the "big blue bump", is primarily
      due to thermal emission from an accretion disk, and that
      variations are induced by a varying X-ray flux irradiating the
      disk. Observations showing that flux variations in the Seyfert 1
      NGC~7469 show progressively longer lags at longer wavelengths
      relative to the UV suggest that the disk radiation is due to
      reprocessed radiation from the inner parts of the disk (Collier
      et al. \cite{col}; Kriss et al. \cite{kri00}). Nandra et al. (\cite{nan}) account
      for the complicated relationship among the time-variable X-ray,
      EUV, and UV fluxes from NGC~7469 by describing the thermal disk
      radiation as a variable seed distribution of soft photons that
      are Compton scattered to create the X-ray flux. These X-rays are
      in turn absorbed and reprocessed by the disk to create the
      observed UV and EUV flux. Since the EUV radiation arises from
      the exponential Wien tail of the flux radiated by the disk,
      slight changes in disk temperature can lead to large variations
      in flux (Marshall et al. \cite{mar97}). In the context of this picture,
      we suggest that the lower amplitude of UV and optical variations
      in a luminous quasar such as HS~1700+6416 is due to its lower
      X-ray to optical luminosity ratio. It is firmly established that
      the X-ray to optical luminosity ratio of AGN is anti-correlated
      with bolometric luminosity (e.g., Kriss \& Canizares
      \cite{kri85}). Since the X-ray radiation in luminous AGN is
      energetically less important, we would expect that X-rays
      illuminating the disks of these objects would play a smaller
      role in determining the radiative output of the disk.

The strong UV variability of luminous QSOs may also have an impact on
both the neighbouring IGM and on the metagalactic EUV background
that ionizes H and HeII.  While for the immediate neighbourhood the
influence of the QSO consists of additional ionization of the
Ly$\alpha$ forest ``clouds'' (Bajtlik, Duncan \& Ostriker, \cite{baj}), this
proximity effect has been proven only statistically with large QSO
samples, since not each QSO shows the expected effect (Bechthold \cite{bec},
Scott et al.  \cite{sco00}). None are shown by HS~1700+6416, one of the most
 luminous QSOs
in the universe where one would expect a strong proximity effect. 
 Among the possible reasons are a finite lifetime of the
present QSO phase insufficient to build up an HII region or a
particularly dense environment in which the QSO resides. The short
term variability that we observed in HS~1700+6416 should have no
observable influence.

\begin{table}\caption{\label{basic} Compilation of
UV and X-ray observations of HS~1700+6416}
\rm
\begin{tabular}{lll}
Satellite/Instrument    &    Date          &   Flux at 1250\\
                        &                  &    [10$^{-15}$ erg s$^{-1}$ cm$^{-2}$ \AA$^{-1}$]\\
\hline
IUE                     &   1988/7/15      &     1.6\\
IUE                     &   1988/11/15     &     2.4\\
                        &        11/17     &        \\
HST/FOS                 &   1992/2/17      &     3.5\\
HST/GHRS                &   1994/7/26      &     1.1\\
HUT                     &   1995/3/4-10    &     1.15\\
HST/GHRS                &   1996/5/29,28   &     1.5\\
HST/STIS                &   1998/7/23-25*  &     1.4\\
HST/STIS                &   2003/5/14,18   &     2.1\\

\hline

&&Flux \\
\hline
FUSE (1050 \AA\,)       &   2002/5/15      &         \\
FUSE (1050 \AA\,)       &   2003/2/27      &    2 $\cdot$ 10$^{-15}$ erg s$^{-1}$ cm$^{-2}$ \AA$^{-1}$    \\
                        &        3/3       &         \\
RASS                    &   1990/12-1991/1 &     0.0115 cts/s\\
ROSAT PSPC              &   1992/11/13     &     0.0088\\
ROSAT PSPC              &   1993/7/22      &     0.0158\\

\end{tabular}
\end{table}

A further aspect is the recent observation of the HeII 304 \AA\,
Ly$\alpha$ forest with FUSE in the lines of sight of HE~2347-4342
(Kriss et al. \cite{kri01}, Shull et al. \cite{shu}) and of HS~1700+6416
(Reimers et al. \cite{rei04}). The column density ratio N(HeII)/N(HI)
= $\eta$, which is roughly proportional to the flux ratio f(911)/f(228)
 of the ionizing background at the corresponding
ionization edges, varies between $\eta \approx $ 1 and $\eta \approx$
several 10$^{2}$ on the scale of 1 Mpc h$^{-1}_{70}$ (Shull et
al. \cite{shu}). This behaviour is not understood and could be a
mixture of radiation transfer effects in the ``cosmic web'' on the
radiation of QSO with a large range of spectral shapes (see
above). Another effect would be the finite lifetime of QSOs (with
light echos of the width of the lifetime in the surrounding medium) or
strong variations of the flux ratios f(911)/f(228) on
even shorter time-scales. While in HS~1700+6416 we have not directly observed
 the corresponding wavelengths (3418 \AA\, and 854 \AA\,), we
estimate from the present observations that
f(911)/f(228) may vary by a factor of 4 in six months.
According to the energy distribution corrected for reddening, Lyman
limit systems and the cumulative effect of the Lyman $\alpha$ forest
shown for the 1994 epoch in Reimers et al. (\cite{rei98}), the
effective $\eta$ caused by HS~1700+6416 varies between 7 (bright phase) and
28 (faint phase).On longer time-scales
(10$^{6}$ yr) this amplitude could be even higher and might be part of
the explanation for small scale variations in the ionizing background.
Ionization and recombination times for HeII/HeI are both in the order 
10$^6$~yr. This implies that, while the ionization equilibrium may not be 
representative of the instantaneous spectrum, short term variations will not
lead to deviations from the ionization equilibrium.

\noindent Acknowledgements:\\
This work has been supported by the Verbundforschung of the BMBF/DLR 
under Grant No. 50 OR 9911. Birgit Fuhrmeister helped with information 
on the epochs of ROSAT observations.

\end{document}